%% file: main.tex
\documentclass[conference]{IEEEtran}
\IEEEoverridecommandlockouts
\usepackage{cite}
\usepackage{amsmath,amssymb,amsfonts}
\usepackage{algorithmic}
\usepackage{graphicx}
\usepackage{textcomp}
\usepackage{xcolor}
\usepackage{mathtools}
\def\BibTeX{{\rm B\kern-.05em{\sc i\kern-.025em b}\kern-.08em
    T\kern-.1667em\lower.7ex\hbox{E}\kern-.125emX}}
\usepackage{array}   
\newcolumntype{L}{>{$}l<{$}} 
\newcommand{\new}[1]{\textcolor{black}{#1}}
\usepackage{ upgreek }
\usepackage{url}
\usepackage{verbatim}
\usepackage{tikz}

\usepackage{moresize}

\renewcommand{\thefootnote}{\alph{footnote}}
\newcommand{\astfootnote}[1]{
\let\oldthefootnote=\thefootnote
\setcounter{footnote}{0}
\renewcommand{\thefootnote}{\fnsymbol{footnote}}
\footnote{#1}
\let\thefootnote=\oldthefootnote
}

\usetikzlibrary{backgrounds,fit,decorations.pathreplacing}                  
\newcommand{\ket}[1]{\ensuremath{\left|#1\right\rangle}} 

\newcommand\mybox[2][]{\tikz[overlay]\node[draw=black,fill=blue!20,inner sep=2pt, anchor=text, rectangle, rounded corners=1mm,#1] {#2};\phantom{#2}}
\newcommand\mygreenbox[2][]{\tikz[overlay]\node[draw=black,fill=green!20,inner sep=2pt, anchor=text, rectangle, rounded corners=1mm,#1] {#2};\phantom{#2}}
\usepackage{qcircuit}

\usepackage{listings}
\usepackage{color}

\definecolor{dkgreen}{rgb}{0,0.6,0}
\definecolor{gray}{rgb}{0.5,0.5,0.5}
\definecolor{mauve}{rgb}{0.58,0,0.82}

\usepackage{booktabs}
\usepackage{amsmath}
\usepackage{etoolbox}

\usepackage{mdframed}

\IEEEpubid{978-1-7281-9418-9/20/\$31.00 \copyright 2020 IEEE}

\newmdtheoremenv{defn}{Definition}
\AtBeginEnvironment{defn}{\small}

\newcommand\myeq{\stackrel{\mathclap{\scriptsize\mbox{def}}}{=}}

\begin{document}

\title{Implementing a Quantum Coin Scheme
\thanks{This publication has emanated from research supported in part by a research grant from Science Foundation Ireland (SFI) and is co-funded under the European Regional Development Fund under Grant
13/RC/2077. Hazel Murray was supported by an Irish Research Council 2017 Government of Ireland Postgraduate Scholarship. Harun Siljak's work was funded by European Union’s Horizon 2020 research and innovation programme under the Marie Skłodowska-Curie grant agreement No~713567.}
}

\author{\IEEEauthorblockN{
Hazel Murray}
\IEEEauthorblockA{\textit{Dept.~Maths \& Stats} /\\ 
\textit{Hamilton Institute} \\
\textit{Maynooth University}\\
Ireland \\
hazel.murray@mu.ie}
\and
\IEEEauthorblockN{
Jerry Horgan}
\IEEEauthorblockA{\textit{TSSG} \\
\textit{Waterford Inst. of Technology}\\
Ireland \\
jhorgan@tssg.org}
\and
\IEEEauthorblockN{
Joao F. Santos}
\IEEEauthorblockA{\textit{CONNECT Centre} \\
\textit{Trinity College Dublin}\\
Ireland \\
facocalj@tcd.ie}
\and
\IEEEauthorblockN{
David Malone}
\IEEEauthorblockA{\textit{Dept.~Maths \& Stats} /\\ 
\textit{Hamilton Institute} \\
\textit{Maynooth University}\\
Ireland \\
david.malone@mu.ie}
\and
\IEEEauthorblockN{
Harun Siljak}
\IEEEauthorblockA{\textit{CONNECT Centre} \\
\textit{Trinity College Dublin}\\
Ireland \\
harun.siljak@tcd.ie}}

\maketitle

\begin{abstract}
Quantum computing has the power to break current cryptographic systems, disrupting online banking, shopping, data storage and communications. Quantum computing also has the power to support stronger more resistant technologies. In this paper, we describe a  digital cash scheme created by Dmitry Gavinsky, which utilises the capability of quantum computing. We contribute by setting out the methods for implementing this scheme. For both the creation and verification of quantum coins we convert the algebraic steps into computing steps. As part of this, we describe the methods used to convert information stored on classical bits to information stored on quantum bits. 
\end{abstract}


\input{contents}

\bibliographystyle{IEEEtran}
\bibliography{IEEEabrv,library}

\end{document}

%% file: contents.tex
\section{Introduction}
Quantum mechanics is the study of the smallest things in nature. At the 1927 Solvay Conference, 29 prominent physicists met to discuss the foundation of today's quantum theory. Amongst the participants were Albert Einstein, Marie Curie, Max Planck, Niels Bohr and Erwin Schrödinger. With their help, an understanding of quantum mechanics has allowed us to develop many modern technologies including MRI scanners, nuclear power, lasers, transistors and semiconductors \cite{miller2008quantum}.

Many years later, in 1980, computation using the principles of quantum mechanics was conceived. Benioff \cite{benioff1980computer} showed that a computer could operate under the laws of quantum mechanics by providing a Schrödinger equation description of Turing machines. In 1988, Yamamoto and Igeta proposed the first physical realization of a quantum computer, it included the quantum equivalent of classical gates \cite{igeta1988quantum}. In 1991, Artur Ekert invented entanglement-based secure communication \cite{ekert1991quantum}. In 1998, a working 2-qubit quantum computer was built by Jones and Mosca at Oxford University \cite{jones1998implementation}. This was the first experimental demonstration of a quantum algorithm. Since then, quantum devices have come a long way. In 2007, Switzerland used quantum technology to secure their voting systems \cite{marks2007quantum}. In Japan, in 2010, a TV conference was secured using quantum key cryptography \cite{sasakione}. China installed a 2000km optical fibre capable of quantum communication, which is being tested for use in banking and communications \cite{qiu2014quantum}. In 2015, a small quantum network was demonstrated by Delft University with plans to build a larger advanced quantum network across the Netherlands \cite{markoff2015sorry}. There are over forty multinational companies investing in quantum computing/communication \cite{vermaas2019quantum}. These include IBM, Google, Microsoft and Intel.
 
Quantum computing has the theoretical power to break certain modern cryptography \cite{mavroeidis2018impact}. In 1994, Peter Shor developed a quantum algorithm that has the power to break some public key cryptographic systems \cite{shor1994algorithms}, such as RSA. In 1996, Grover's algorithm was developed, which reduced the effectiveness of symmetric key cryptographic systems \cite{grover1996algorithms}. Without cryptography, much of our online banking, shopping and data storage technology would no longer be usable. 

Though quantum computing has the power to break some of our current systems, it also holds the key to unlocking solutions that exceed the bounds of our current computational capabilities. Quantum technology has particularly useful qualities for applications to communication systems, privacy and security. In 2019, RIPE NCC \cite{ripe} ran the first Pan-European Quantum Internet Hackathon. This event connected experts from six different locations and tasked them with solving open problems and developing technical infrastructure to allow the evolution of the Quantum Internet. Among other developments, teams successfully worked on Device-Independent Quantum Key Distribution, a Quantum version of Byzantine Agreement, Quantum Key Distribution in OpenSSL, Quantum-Cheque Protocol, Quantum Anonymous Transmission, Entanglement Routing and Quantum VPN. For more details on these projects see the Github repository \cite{github_rep_hackathon}.

\IEEEpubidadjcol
This paper arose from work completed at the Irish node of the Pan-European Quantum Internet Hackathon \cite{connect_hosts_hackathon}. Our goal was to develop the implementation steps necessary for a digital cash protocol based on quantum technologies; denoted a Quantum Coin Scheme. In this paper, we introduce quantum mechanics and describe its relevance to applications in banking and communication systems. We describe the mechanisms involved in creating and manipulating quantum bits. Finally, we describe contributions that allow for the implementation of Gavinsky's \cite{gavinsky2012quantum} theoretical quantum coin protocol.

In Sec.~\ref{quantum-basics}, we explain the underlying properties of quantum mechanics that make it valuable for communication and computation technologies. In Sec.~\ref{q-money-background}, we describe related work and the development of quantum money. Sec.~\ref{notation,terminolgy} introduces the notation and terminology used in this paper. In Sec.~\ref{sec:creation}, \new{we describe one definition of a quantum coin (denoted a $\mathcal{Q}$-coin) and demonstrate the steps necessary for creating it. This involves the creation of a method for converting classical bits to quantum information. This is used to show how to create quantum coins for use in quantum money transactions. Sec.~\ref{sec:veri} details the processes necessary for using these $\mathcal{Q}$-coins in the implementing of Gavinsky's quantum coin validation. Sec.~\ref{sec:summary} summarises some feastures of the scheme.}

\section{Background}
\subsection{The Power Of Quantum Computing}\label{quantum-basics}
Quantum mechanics is interesting because it contains properties that are at odds with our general understanding of classical physics. Here we will give a brief overview of the properties we utilise. Many more detailed descriptions are available (e.g. \cite{miller2008quantum}).

Used for information storage, a classical bit can take the value 0 or 1. A qubit is the quantum equivalent to a classical bit. 
Qubits have three important properties that makes them fascinating as an alternative to our classical view of information: \textit{superposition}, \textit{measurement} and \textit{entanglement}.
 
 The first property, \textit{superposition}, describes the fact that a qubit can take the value of both 0 and 1 at the same time! Imagine we have two classical bits, these can represent 4 states: either both bits are zero: 00, one bit is zero and the other is one: 01 or 10, or both bits are one: 11. If we have 2 qubits, we can still represent these 4 states: 00, 01, 10 and 11. However, because of superposition, the 2 qubits can represent a mix of all 4 states at the same time. This gives quantum computers the capacity to complete computations in parallel and where $n$ classical bits allow $n$ computations, $n$ qubits can allow $2^n$ computations.

The second property is \textit{measurement}. In classical mechanics, looking at something does not change its state. In quantum mechanics, a qubit can be in a superposition of both 0 and 1 at the same time and when measured it must \textit{collapse} to either 0 or 1. The state of a quantum bit is represented by a wave function, where $\ket{0}$ is the 0 wave function, $\ket{1}$ is the 1 wave function, and $\alpha\ket{0}+\beta\ket{1}$ is a superposition. A wave function that is composed of only $\ket{0}$ or $\ket{1}$ is called an \emph{eigenstate}. If we measure a wave function to see if it is a 0 or a 1, then there is a probability $|\alpha|^2$ of measuring 0 and $|\beta|^2$ of 1. Naturally, we need to normalise so that $|\alpha|^2 + |\beta|^2 = 1$. Each qubit can be represented as a wave function and on measurement of the wave function as 0 or 1 it collapses and becomes $\ket{0}$ or $\ket{1}$. This has implications for security. If we send classical bits from one place to another, we have no way to know whether they were observed by a malicious user. However, if we communicate using qubits, a malicious user who observes the qubits will collapse the wave function and we will know that the message was intercepted. 

One interesting thing to note, is that we are collapsing the wave function for the property we are measuring, this is called the \textit{basis} that we are measuring with respect to. Imagine there are two measurements on a qubit, say its position can be $A$ or $B$ and its momentum can be 0 or 1. We measure its position and the wave function collapses to $\ket{A}$. If we continue to consecutively measure with respect to the position basis then we will continue to get A. If we then measure using the momentum basis, the momentum wave function collapses to $\ket{1}$, and the position variable is again probabilistic. So if we remeasure the position it will return either $A$ or $B$ with some probability. It is true for any measurable qubit attributes. This is an example of the famous Heisenberg Uncertainty Principle \cite{heisenberg1985anschaulichen}.

The third and, according the Einstein, the `spooky' property of quantum mechanics is \textit{entanglement}. Take two qubits that are entangled and let us move them to opposite ends of the globe. If we measure one of the qubits then we know that we will get the same measurement for the second, entangled, qubit. Imagine we take the first qubit and measure it using a momentum basis and get 1. Then the other qubit will also measure as 1. This is remarkable since each returned result is a function of probabilities $|\alpha|^2$ and $|\beta|^2$. This relationship gives us the ability to send information via these two entangled qubits (but not faster-than-light, as we might be tempted to attempt \cite{bruss2000approximate}).

These properties have applications in our computing and communications infrastructure. We are going to look at the applications of qubits to our online representation of coins that are used to transfer funds between bank accounts.

\subsection{Quantum Money}\label{q-money-background}
In classical cryptography the concept of digital cash has been well-explored \cite{chaum1988untraceable}. Let us briefly describe a classical digital cash scheme.

Every coin issued by the bank is represented by a secret string $s$. These strings are known to the bank and to the current coin holder (Alice). Suppose Alice wishes to pay Bob, she will want to pass her coin to Bob:
\begin{itemize}
    \item Alice sends her string $s$ to the Bank and tells the bank she wants to send the coin to Bob, 
    \item The bank checks if the string sent by Alice is valid. If so, the bank erases the string $s$ from the list of valid strings and adds a newly generated secret string $s'$ to the list.
    \item The bank sends $s'$ to Bob; henceforth, Bob holds the coin.
\end{itemize}

For classical digital cash schemes, the main concern is the double-spending problem, where a user spends the same digital coin multiple times. One solution, as above, is to include a verification of each token with a bank. However, an intruder who pretends to be the bank can steal a valid coin from its fair holder who wants it to be verified. 

\begin{defn}[Coin]
A coin is a unique object that can be created by a trusted mint (or bank) and then circulated among untrusted holders. 
\end{defn}

For quantum money we will also need our coins to be non-counterfeitable. Conveniently, qubits have a property described as the \textit{no-cloning theorem} \cite{wooters1982quantum} that makes them perfect to be applied to quantum money. The no cloning theorem tells us that it is impossible to create an identical copy of a quantum state. Wiesner argued that this property allows us to create quantum coins that are unforgeable, something that is impossible with our classical physical money. In 1983, Wiesner \cite{wiesner1983conjugate} and Bennett, Brassard, Breidbard, and Wiesner \cite{bennett1983quantum} conceived the first quantum money schemes. 

In 2003, Tokunaga, Okamoto, and Imoto give a scheme for non-transferable anonymous quantum cash with online verification \cite{tokunaga2003anonymous}. In 2010, Mosca and Stebila present a new type of quantum money which they call quantum coins \cite{mosca2010quantum}. These coins are transferable, locally verifiable, and unforgeable, and have some anonymity properties. However both these schemes require quantum communication with a bank and are also both susceptible to an adaptive attack conceived by Lutomirski \cite{lutomirski2010online}.

In 2012, Gavinsky proposed a new quantum coin scheme that allows classical verification of coins. In the version of a quantum internet where quantum and classical computers will work in synchrony this is an ideal scheme. We can leverage the power of quantum bits without the requirement for every user to possess quantum communication technology. Gavinsky's scheme is secure against adaptive adversaries, the coins are exponentially hard to counterfeit, verification can be conducted via insecure communication lines, the bank's database is static and can therefore be decentralized, and the scheme protects against a malicious user masquerading as a bank. The coins are limited to a certain number of verifications, which trade off against the size of the coin (number qubits). However, Gavinsky shows that this dependency is optimal.

In this paper we outline the methods necessary for implementing Gavinsky's quantum coin scheme. We specifically describe the physical gates necessary for the creation and verification of the quantum coins.

\subsection{Notation}\label{notation,terminolgy}
This section introduces the notation used in the paper. 
\subsubsection{Matrix representation of qubits}
In Sec.~\ref{quantum-basics}, we explained that a qubit can be in a superposition of both 0 and 1 and is represented as the vector:
\[
    q = \alpha\ket{0}+\beta\ket{1},
\]
where states $\ket{0}$ and $\ket{1}$ form a basis for the vector space and $\alpha$ and $\beta$ are complex numbers that indicate the amplitude of the state. The amplitude squared tells us the probability of the state occurring. The above vector describes one qubit that can be in a superposition of two states. In this paper we are generally working with 2 qubits, which have 4 possible states. We call these states $\ket{00}$, $\ket{01}$, $\ket{10}$, $\ket{11}$. These states can also be descibed in matrix form as: 
\[
    \ket{00} = \begin{bmatrix} 1 \\ 0 \\ 0 \\ 0\end{bmatrix}, \ket{01} = \begin{bmatrix} 0 \\ 1 \\ 0 \\ 0\end{bmatrix}, \ket{10} = \begin{bmatrix} 0 \\ 0 \\ 1 \\ 0\end{bmatrix} \mbox{ and }\ket{11} = \begin{bmatrix} 0 \\ 0 \\ 0 \\ 1\end{bmatrix}.
\]
The four states together form the \textit{basis}. Each state has a certain probability of occurring, determined by the amplitudes $\alpha, \beta, \gamma$ and $\delta$ of the wave function. The basis matrix times the amplitude vector gives us the wave function for our two qubits: 
\[
     \setlength{\arraycolsep}{3pt}{\begin{bmatrix}\ket{00}&\ket{01}&\ket{10}&\ket{11} \end{bmatrix}} \begin{bmatrix} \alpha \\ \beta \\ \gamma \\ \delta\end{bmatrix} = \alpha\ket{00} + \beta\ket{01} +\gamma\ket{10} + \delta\ket{11}.
\]
For simplicity, given the context of the basis, we can just report the amplitude matrix when describing the qubit pair. We use the subscript $A$ to denote an amplitude matrix: $\begin{bmatrix} \alpha & \beta & \gamma & \delta \end{bmatrix}_A$.

\subsubsection{Quantum gates}
Both classical and quantum logic gates take binary inputs and produce a single binary output. Quantum gates, like classical gates, can be combined into a circuit. One benefit of quantum gates is that, unlike classical gates, they are always reversible. This means that no information is lost when qubits travel through quantum gates. 

In this paper we will use six quantum gates. In Tab.~\ref{tab:gates}, we define each gate by stating the gate's function, symbol and matrix representation. To learn more about quantum gates see \cite{nielsen2002quantum}.

\begin{table}
    \caption{Overview of six quantum gates}
    \label{tab:gates}
    \centering
    \begin{tabular}{c|c|c|p{2.5cm}}
    Name& Gate & Matrix &Description\\\hline
     Hadamard& \Qcircuit @C=1em @R=.7em {& \gate{H} & \qw} &  $\frac{1}{\sqrt{2}}\begin{bmatrix}1 & 1 \\1 &-1 \end{bmatrix}$& \vspace{-1em}Maps \ket{0} to $\frac{\ket{0} + \ket{1}}{\sqrt{2}}$ and \ket{1} to $\frac{\ket{0} - \ket{1}}{\sqrt{2}}$. It sends a qubit into a superposition.\\\hline
     Pauli-X & \Qcircuit @C=1em @R=.7em {& \gate{X} & \qw} & $\begin{bmatrix}0& 1 \\1 &0 \end{bmatrix}$& \vspace{-1em}Maps \ket{0} to \ket{1} and vice versa. Equivalent of the classical NOT gate.\\\hline
     Pauli-Z & \Qcircuit @C=1em @R=.7em {& \gate{Z} & \qw} & $\begin{bmatrix}1& 0 \\0 &-1 \end{bmatrix}$ & \vspace{-1em}It leaves the basis state \ket{0} unchanged and maps \ket{1} to -\ket{1}. It is sometimes called phase-flip.\\\hline
     CNOT & \Qcircuit @C=1em @R=1em {& \ctrl{1} & \qw \\& \targ  & \qw}&  $\begin{bmatrix}1& 0&0&0 \\0&1&0&0\\ 0&0&0&1\\ 0&0&1&0 \end{bmatrix}$ & \vspace{-2em}Flips the second qubit (the target qubit) if and only if the first qubit is \ket{1}. The CNOT gate allows us to entangle two input qubits.\\\hline
     SWAP & \Qcircuit @C=1em @R=1em {& \qswap & \qw\\& \qswap \qwx & \qw} & $\begin{bmatrix}1& 0 &0&0\\0&0&1&0\\0&1&0&0\\0&0&0&1 \end{bmatrix}$ & \vspace{-2em} The swap gate swaps two qubits.\\\hline
     Identity & \Qcircuit @C=1em @R=.7em {& \gate{I} & \qw} & $\begin{bmatrix}1 & 0 \\0 &1 \end{bmatrix}$ & \vspace{-1em} Leaves the basis states \ket{0} and \ket{1} unchanged. It can be used to expand gates so that they can work on multiple qubits. \\\hline
    \end{tabular}
\end{table}




\section{Creation of $\mathcal{Q}$ coins}\label{sec:creation}
Let us describe Gavinsky's definition of a coin, named $\mathcal{Q}$-coin, then we will describe how it is created.
\begin{defn}[$\mathcal{Q}$-coins]\label{def:Qcoin}
For each coin, a bank holds a secret record consisting of $k$ entries $x_1, \dots, x_k$ s.t. $x_i \in \{0,1\}^4$ (i.e., the secret record contains $k$ strings of $4$ classical bits). 

A ``fresh'' $\mathcal{Q}$-coin is then created corresponding to this record $(x_1,\dots, x_k)$. \textbf{The coin consists of:}
\begin{itemize}
    \item $k$ quantum registers consisting of 2 qubits each, where the $i$'th register contains a specific state $\ket{\alpha(x_i)}$;
    \item a $k$-bit classical register $\mathcal{P}$. This consists of $k$ binary markers that indicate whether the $i$'th quantum register has been used in previous validation processes. The values of $\mathcal{P}$ are initially set to $0^k$;
    \item a unique identification number.
\end{itemize}
\end{defn}

Creation of the coins requires the conversion of the four classical bits to two quantum bits. Algebraically, we use the formula below for conversion. Because this conversion satisfies the 4-bit version of the Hidden Matching Problem (HMP) \cite{bar2004exponential}, Gavinsky calls these quantum registers  $\textit{HMP}_4$-states. 



\begin{defn}[$\textit{HMP}_4$-states]\label{def:HMP-states}
    Let $x \in \{0,1\}^4$. The corresponding $\textit{HMP}_4$-\textit{state} is 
    \[
        \ket{\alpha(x)} \myeq \frac{1}{\sqrt{4}} \sum_{1 \leq i \leq 4} (-1)^{x_i} \ket{(i-1)_2},
    \]
    where $(\cdot)_2$ denotes writing a number in base 2.
\end{defn}

For example, the 4 classical bits $x = 0110$ are converted to the state $\ket{\alpha(0110)}$, which is
  \begin{align*}
     \frac{1}{\sqrt{4}} \huge((-1)^{0} \ket{00} + (-1)^{1} \ket{01} + (-1)^{1} \ket{10} + (-1)^{0} \ket{11}\huge) \\
     = \frac{1}{2} \huge(\ket{00}  -\ket{01} - \ket{10}  +\ket{11}\huge).
\end{align*}%
 Up to normalisation, this can be represented by the following amplitude matrix $\begin{bmatrix} 1 & -1 & -1 & \hphantom{+}1\end{bmatrix}_A$.

\subsection{Implementation}\label{sec:create}

$\mathcal{Q}$-coins require a quantum representation of 4 bit classical strings. There are 16 possible combinations of 4 bits. Each of these needs to be uniquely represented by a quantum register according to the conversion specified in Def.~\ref{def:HMP-states}. In this section, we show how to prepare these 16 $\textit{HMP}_4$-states.

\subsubsection{No entanglement}
Given any two quantum bits. Let $q_1 = \alpha\ket{0}_1 +\beta\ket{1}_1$ and $q_2 = \gamma\ket{0}_2 +\delta\ket{1}_2$, where the subscript denotes the qubit the state belongs to. The state space of a composite systems is the tensor product of the state spaces of the components, so for our two qubits
    \begin{align*}
        q_1 \otimes q_2 &= (\alpha\ket{0}_1 +\beta\ket{1}_1)(\gamma\ket{0}_2 +\delta\ket{1}_2)\\ &= \alpha\gamma\ket{00} + \alpha\delta\ket{01} + \beta\gamma\ket{10} + \beta\delta\ket{11}\\
        &\equiv \begin{bmatrix} \alpha\gamma & \alpha\delta & \beta\gamma & \beta\delta\end{bmatrix}_A
    \end{align*} 
For simple input states, $\alpha, \beta,\gamma$ and $\delta$ can each take either -1 or +1. Thus, by manipulating the state of the initial qubits, $q_1$ and $q_2$, we can create the following state spaces:

    \begin{center}
    \small{
    \setlength{\tabcolsep}{3pt}{
    \begin{tabular}{L L L L}
        \alpha=-1 & \beta,\gamma,\delta=1 & \rightarrow & \begin{bmatrix} -1 & -1 & \hphantom{+}1 & \hphantom{+}1\end{bmatrix}_A \hphantom{l} \myeq \hphantom{l}  \ket{Q_1}\\
        \beta=-1 & \alpha,\gamma,\delta=1 & \rightarrow & \begin{bmatrix} \hphantom{+}1 & \hphantom{+}1 & -1 & -1\end{bmatrix}_A \hphantom{l} \myeq \hphantom{l}\ket{Q_2}\\
        \gamma=-1 & \alpha,\beta,\delta=1 & \rightarrow & \begin{bmatrix} -1 & \hphantom{+}1 & -1 & \hphantom{+}1\end{bmatrix}_A \hphantom{l} \myeq \hphantom{l}\ket{Q_3}\\
        \delta=-1 & \alpha,\beta,\gamma=1 & \rightarrow & \begin{bmatrix} \hphantom{+}1 & -1 & \hphantom{+}1 & -1\end{bmatrix}_A \hphantom{l} \myeq \hphantom{l}\ket{Q_4}\\
        \alpha,\beta=-1 & \gamma,\delta=1 & \rightarrow & \begin{bmatrix} -1 & -1 & -1 & -1\end{bmatrix}_A \hphantom{l} \myeq \hphantom{l}\ket{Q_5}\\
        \alpha,\gamma=-1 & \beta,\delta=1 & \rightarrow & \begin{bmatrix} \hphantom{+}1 & -1 & -1 & \hphantom{+}1\end{bmatrix}_A \hphantom{l} \myeq \hphantom{l}\ket{Q_6}\\
        \alpha,\delta=-1 & \beta,\gamma=1 & \rightarrow & \begin{bmatrix} -1 & \hphantom{+}1 & \hphantom{+}1 & -1\end{bmatrix}_A \hphantom{l} \myeq \hphantom{l}\ket{Q_7}\\
        \multicolumn{2}{L}{\alpha,\beta,\gamma,\delta=-1} &\rightarrow & \begin{bmatrix} \hphantom{+}1 & \hphantom{+}1 & \hphantom{+}1 & \hphantom{+}1\end{bmatrix}_A \hphantom{l} \myeq \hphantom{l}\ket{Q_8}.\\
    \end{tabular}}}
    \end{center}
All other combinations give repetitions of these 8 $\textit{HMP}_4$-states, so we use entanglement to create the other states.

\subsubsection{With entanglement}
To convert the remaining 8 classical strings, we begin by entangling the 2 input qubits, $q_1$ and $q_2$. We then send these entangled qubits through gates to manipulate them to create the 8 required quantum registers. 

We create entangled states by taking simple inputs and putting them through a Hadamard and a CNOT gate. The Hadamard gate is applied to the first qubit and sends it into a superposition. Then the CNOT gate is applied to both qubits. This conditional gate creates an entanglement between the two qubits. See Tab.~\ref{tab:gates} for the matrix description of both gates. 
 
 By specifying four different initial states of the qubits, $q_1$ and $q_2$, we can create four different entangled states. These are called \emph{Bell states}. 
 
     As an example, by starting both qubits in the eigenstate \ket{0} we create the first Bell state, called \ket{\Phi^+}:
    
\input{diagram}

    \begin{defn}[Bell States]
   The Bell states are four specific maximally entangled quantum states of two qubits given by:
   \begin{align*}
       &\ket{\Phi^+}= \frac{1}{\sqrt{2}}(\ket{00} + \ket{11}) \equiv  \begin{bmatrix} 1&0&0&1\end{bmatrix}^T\\
       & \ket{\Phi^-}= \frac{1}{\sqrt{2}}(\ket{00} - \ket{11}) \equiv \begin{bmatrix} 1&0&0&-1\end{bmatrix}^T\\
       &\ket{\Psi^+}= \frac{1}{\sqrt{2}}(\ket{01} + \ket{10}) \equiv \begin{bmatrix} 0&1&1&0\end{bmatrix}^T\\
       &\ket{\Psi^-}= \frac{1}{\sqrt{2}}(\ket{01} - \ket{10}) \equiv \begin{bmatrix} 0&1&-1&0\end{bmatrix}^T
   \end{align*}
    \end{defn}

Using the Bell states we can generate the required remaining 8 combinations by using a sequence of quantum gates. 

For example, if we create the Bell state \ket{\Phi^+} and pass it through an extended Hadamard gate, $(H \otimes I)$, we can create a ninth $\textit{HMP}_4$-state; $\begin{bmatrix}1 & \hphantom{+}1 & \hphantom{+}1 & -1\end{bmatrix}_A$:
    \input{diagram2.tex}

Below we describe the input Bell state and the combination of gates used to create the last 8 linear combinations. $I$ denotes the identity matrix, $H$ denotes the Hadamard gate, $\otimes$ denotes the tensor product (gates wired in parallel) and $\times$ denotes the ordinary matrix cross product (serially wired gates).
 
    \begin{center}
        \setlength{\tabcolsep}{3pt}{
        \begin{tabular}{L L L}
             (H \otimes I) \times \ket{\Phi^+}&  \rightarrow &  \frac{1}{2}\begin{bmatrix} \hphantom{+}1 & \hphantom{+}1 & \hphantom{+}1 & -1\end{bmatrix}_A \hphantom{l} \myeq \hphantom{l} \ket{Q_9}\\
             (H \otimes I) \times \ket{\Psi^+}&  \rightarrow &  \frac{1}{2}\begin{bmatrix} \hphantom{+}1 & \hphantom{+}1 & -1 & \hphantom{+}1\end{bmatrix}_A  \hphantom{l} \myeq \hphantom{l} \ket{Q_{10}}\\
             (H \otimes I) \times \ket{\Phi^-}&  \rightarrow &  \frac{1}{2}\begin{bmatrix} \hphantom{+}1 & -1 & \hphantom{+}1 & \hphantom{+}1\end{bmatrix}_A \hphantom{l} \myeq \hphantom{l}\ket{Q_{11}}\\
             (X \otimes I) \times \ket{Q_{10}} & \rightarrow &  \frac{1}{2}\begin{bmatrix} -1 & \hphantom{+}1 & \hphantom{+}1 & \hphantom{+}1\end{bmatrix}_A \hphantom{l} \myeq \hphantom{l}\ket{Q_{12}}\\
             (Z \otimes I) \times \ket{Q_{11}}& \rightarrow & \frac{1}{2}\begin{bmatrix} \hphantom{+}1 & -1 & -1 & -1\end{bmatrix}_A \hphantom{l} \myeq \hphantom{l}\ket{Q_{13}}\\
             (Z \otimes I) \times \ket{Q_{12}}& \rightarrow & \frac{1}{2}\begin{bmatrix} -1 & \hphantom{+}1 & -1 & -1\end{bmatrix}_A \hphantom{l} \myeq \hphantom{l}\ket{Q_{14}}\\
            (I \otimes Z) \times \ket{Q_{12}}& \rightarrow & \frac{1}{2}\begin{bmatrix} -1 & -1 & \hphantom{+}1 & -1\end{bmatrix}_A \hphantom{l} \myeq \hphantom{l}\ket{Q_{15}}\\
            (X \otimes I) \times \ket{Q_{14}}& \rightarrow & \frac{1}{2}\begin{bmatrix} -1 & -1 & -1 & \hphantom{+}1\end{bmatrix}_A \hphantom{l} \myeq \hphantom{l}\ket{Q_{16}}\\
            %
        \end{tabular}}
    \end{center}
 
 We have shown how to create 16 quantum registers which uniquely represent the 16 classical combinations of 4 bits. For creation of the quantum coin, the mapping of classical bits to quantum bits can be hard-coded and only needs to be completed once. As described in \cite{garcia2011equivalent}, there will be alternative circuit configurations that will produce equivalent results.

 
\section{Verification}\label{sec:veri}
Let us now introduce Gavinsky's coin verification protocol. The protocol involves the key holder proving that they hold the coin, without needing to reveal its full details to the bank. It is a version of a zero knowledge protocol.

As described in Def.~\ref{def:Qcoin}, coins with unique identification numbers have been created by the bank. The bank holds a secret record $(x_1, \dots x_k)$ associated with the identification number for each one of its created coins. A coin holder Bob has one such $\mathcal{Q}$-coin which contains the following information:

 \begin{center}
    \begin{tabular}{|c|c|}\hline
    \multicolumn{2}{|c|}{Identification number}\\\hline
        \hphantom{hi}$\mathcal{P}_1$\hphantom{hi} & $\ket{\alpha(x_1)}$  \\\hline
        $\mathcal{P}_2$ & $\ket{\alpha(x_2)}$\\\hline
        $\vdots$ & $\vdots$\\\hline
        $\mathcal{P}_k$ & $\ket{\alpha(x_k)}$\\\hline
    \end{tabular}
     \end{center}
    $\mathcal{P}_i$ marks whether the quantum register, \ket{\alpha(x_i)}, at position $i$ has previously been used for verification. The coin holder, Bob, wishes to verify the coin's authenticity.

The verification is based on the communication complexity problem called the Hidden Matching Problem introduced by Bar-Yossef et al. \cite{bar2004exponential}. The Hidden Matching Problem (\textit{HMP}) is defined as follows: 

\begin{defn}[$\textit{HMP}_4$  condition]\label{def:HMP}
For $x \in \{0,1\}^4$ and $m,a,b \in \{0,1\}$, we say that
    $(x,m,a,b) \in \textit{HMP}_4$ if \[
    b= 
\begin{cases}
    x_1 \otimes x_{2+m} & \text{if } a= 0\\
    x_{3-m} \otimes x_4 & \text{if } a=1
\end{cases}
\]
\end{defn}
In the below protocol, Bob will provide  values $(a_i,b_i)$ to the bank. The bank holds the values $x_i$ (the classical bit strings) and $m_i$. If $\forall i \; (x_i,m_i,a_i,b_i)\in \textit{HMP}_4$, then the bank can verify that Bob does in fact hold the $\mathcal{Q}$-coin corresponding to the classical values $x_i$. We will now describe the specific steps involved and the methods for implementation.

\subsection{Steps in Gavinsky's protocol}\label{gavinsky}
\begin{center}
\small
\begin{tabular}{lr}\hline
    Coin Holder &Bank\\\hline
    \multicolumn{2}{p{8.1cm}}{Step 1: \textcolor{red}{The holder sends the identification number on the coin they hold to the bank.}}\\
     \mybox{ID number} $\longrightarrow$&\\\hline
     \multicolumn{2}{p{8.1cm}}{Step 2: \textcolor{red}{The bank uses the identification number to look up the secret record of $k$ classical strings, $(x_1,\dots,x_k)$, which were created for this coin.}}\\
     \multicolumn{2}{p{8.1cm}}{\textcolor{red}{The bank chooses $t$ indexes (s.t. $3|t$ \& $t\leq k$) at random between 1 and $k$ and sends them to the coin holder:}}\\
     &$\mathcal{L}_{bank} \subset [k]$,\\
     &s.t. $|\mathcal{L}_{bank}| = t$ and $3|t$\\
     &$\longleftarrow$ \mybox{$\mathcal{L}_{bank}$}\smallskip\\\hline
     \multicolumn{2}{p{8.1cm}}{Step 3: \textcolor{red}{The holder randomly selects $2t/3$ of the values sent by the bank that have not been used for validation before, i.e. $\mathcal{P}_i = 0$:}}\\
     $\mathcal{L}_{holder}\subset \mathcal{L}_{bank},$&\\
     s.t. $\forall i \in \mathcal{L}_{holder}\; P_i = 0$&\\
     and $|\mathcal{L}_{holder}| = 2t/3$.\\ 
     \multicolumn{2}{p{8.1cm}}{\textcolor{red}{The holder sends $\mathcal{L}_{holder}$ to the bank and marks those elements as used in the register $\mathcal{P}$:}}\\
     $\mathcal{P}_i = 1 \;\forall i \in \mathcal{L}_{holder}$&\\
     \mybox{$\mathcal{L}_{holder}$} $\longrightarrow$&\smallskip\\\hline
     \multicolumn{2}{p{8.1cm}}{Step 4: \textcolor{red}{For each index in $\mathcal{L}_{holder}$, the bank randomly chooses an $m$ equal to 0 or 1 and sends these back to the coin holder:}}\\
     &$\forall i \in \mathcal{L}_{holder}$\\
     &$m_i \in \{0,1\}$\\
     &$\longleftarrow$ \mybox{$m_i$}\smallskip\\\hline
     \multicolumn{2}{p{8.1cm}}{Step 5: \textcolor{red}{The holder measures the quantum registers, $\ket{\alpha(x_i)}, \forall i \in \mathcal{L}_{holder}$. The basis used for the measurement is determined by the value $m$ sent by the bank:}}\\
    $\forall i \in \mathcal{L}_{holder}$&\\
    measure $\ket{\alpha(x_i)} \Rightarrow (a_i,b_i)$.\\
    \multicolumn{2}{p{8.1cm}}{\textcolor{red}{The coin holder sends the output values $(a_i,b_i)$ corresponding to each $i \in \mathcal{L}_{holder}$ to the bank:}}\smallskip\\
    \mybox{$(a_i,b_i)$}$\longrightarrow$&\smallskip\\\hline
    \multicolumn{2}{p{8.1cm}}{Step 6: \textcolor{red}{The bank checks whether $(x_i, m_i, a_i, b_i) \in \textit{HMP}_4$ for all $i \in \mathcal{L}_{holder}$ (by Def.~\ref{def:HMP}):}}\\
    &if $(x_i, m_i, a_i, b_i) \in \textit{HMP}_4$\\
    & $\forall i \in \mathcal{L}_{holder}$\\
    &$\longleftarrow$\mygreenbox{Coin is Valid}\\\hline
\end{tabular}
\end{center}
\normalsize

Note that a bank produces \textit{fresh} $\mathcal{Q}$-coins but as a $\mathcal{Q}$-coin goes through more and more verification protocols, its quantum registers lose their original content; when a quantum state/register is measured it collapses. For each quantum verification we measure $2t/3$ quantum registers. Hence, we collapse $2t/3$ registers every time we verify the coin's identity. Depending on the level of trust we require and how long we want the coin to last we can choose the value $t$.

We will expand on the steps in the above protocol and translate them into computational steps. We begin by describing the implementation of Steps 1 to 4 in Sec.~\ref{sec:1-4}, the implementation of Step 5 in Sec.~\ref{sec:5}, and finally we describe Step 6 in Sec.~\ref{sec:6}. 

\subsection{Implementation of Steps 1--4}\label{sec:1-4}

Steps 1--4 are classical steps. They involve classical correspondence between Bob and the bank. Below we include the code for both parties. The code is written for SimulaQron \cite{dahlberg2018simulaqron}. SimulaQron is a free quantum internet simulator. It allows users to program their your own quantum internet applications.

\lstset{language=Python}
\begin{lstlisting}[title=\normalsize{Code for bank --- Alice},label=lst:bank]
def verify_coin(register,t):
    #Step 2
    register_c = list(register)
    m_s = []
    list_of_random_indexes = random.sample(register_c, t)
    with CQCConnection("Alice") as bank:
        # Step 2 cont; send the list of indexes to the coin holder
        bank.sendClassical("Bob", list_of_random_indexes)
        # Wait for receiver to send back subset of list
        #Receive output from step 3
        index_list = bank.recvClassical()
        rlist = list(index_list)
        # Step 4; send randomly either 0 or 1 to correspond to each index in Bob's list. 
        for c,i in enumerate(rlist):
            register_c.remove(i)
            m_s.append(random.randint(0,1))
        bank.sendClassical("Bob", m_s)
\end{lstlisting}

\begin{lstlisting}[title=\normalsize{Code for coin holder --- Bob},label=lst:holder]
def verify_coin():
    print("Verify Coin ID 1")
    with CQCConnection("Bob") as Bob:
        register_c = list(range(8))
        #Receive output from step 2
        list_of_random_indexes = Bob.recvClassical()
        #Step 3; chooses a subset of the index list received of size 2t/3
        t = len(list_of_random_indexes)
        local_selection = random.sample(list_of_random_indexes, 2*t/3)
        #Step 3 cont; send subset to bank and mark those indexes as used. 
        Bob.sendClassical("Alice", local_selection)
        for c,i in enumerate(local_selection):
            register_c.remove(i)
        #Receive output from step 4
        m_s = Bob.recvClassical()
\end{lstlisting}

Additional code for our SimulaQron programs can be found on Github in files \verb|bank.py| and \verb|bob.py| \cite{github_rep_qcoin}.

\subsection{Step 5}\label{sec:5}
The holder receives the values $m$ corresponding to each element of $\mathcal{L}_{holder}$. The value $m$ is used to specify the basis, $\{v_1,v_2,v_3,v_4\}$, which each quantum register (with indexes $i \in  \mathcal{L}_{holder}$) will be measured with respect to. 

\subsubsection{Method}

Formally $m$ is defined as an $\textit{HMP}_4$-query. The bank queries Bob for the measurements of the $i$ quantum registers with respect to specific $m$ values. The outputted values will satisfy the $\textit{HMP}_4$ condition.

It is known that $(x,m,a,b) \in \textit{HMP}_4$ always \cite{bar2004exponential}.



\begin{defn}[$\textit{HMP}_4$-queries]
 \label{def:HMP-queries}
\small{
    If $m = 0$, let
    \ssmall{
    \begin{align*}
        v_1 \myeq \frac{\ket{00} + \ket{01}}{\sqrt{2}}, v_2 \myeq \frac{\ket{00} - \ket{01}}{\sqrt{2}}, v_3 \myeq \frac{\ket{10} + \ket{11}}{\sqrt{2}}, v_4 \myeq \frac{\ket{10} - \ket{11}}{\sqrt{2}}
    \end{align*}}\small
    otherwise if $m=1$, let 
    \ssmall{
    \begin{align*}
        v_1 \myeq \frac{\ket{00} + \ket{10}}{\sqrt{2}}, v_2 \myeq \frac{\ket{00} - \ket{10}}{\sqrt{2}}, v_3 \myeq \frac{\ket{01} + \ket{11}}{\sqrt{2}}, v_4 \myeq \frac{\ket{01} - \ket{11}}{\sqrt{2}}
    \end{align*}}\small
    Measure \ket{\alpha(x)} in the basis $\{v_1,v_2,v_3,v_4\}$. Let 
 \[
    (a,b)= 
\begin{cases}
    (0,0) ,& \text{if } v_1\\
    (0,1) ,& \text{if } v_2\\
    (1,0) ,& \text{if } v_3\\
    (1,1) ,& \text{if } v_4.\\
\end{cases}
\]}
\end{defn}

\subsubsection{Implementation}
Step 5 involves generating an output $(a,b)$ determined by the outcome of measuring with the basis determined by $m$. 
As per Sec.~\ref{quantum-basics}, the basis describes what property of the qubit we are measuring and specifically what gates the qubits are passed through. The values $\{v_1,v_2,v_3,v_4\}$ specify the column vectors that make up our basis matrix.

For example, if $m=0$ then Bob is told to use the following vectors to form the basis matrix: 
\[
v_1=\begin{bmatrix}1 \\1 \\0\\0\end{bmatrix}, v_2=\begin{bmatrix}1 \\-1 \\0\\0\end{bmatrix}, v_3=\begin{bmatrix}0\\0\\1\\1\end{bmatrix},v_4=\begin{bmatrix}0\\0\\1\\-1\end{bmatrix}
\]

The matrix formed by these vectors is equivalent to an expanded Hadamard gate:

\[I \otimes H = \frac{1}{\sqrt{2}}\begin{bmatrix}1& \hphantom{+}1 &\hphantom{+}0 &\hphantom{+}0\\1 &-1 &\hphantom{+}0 &\hphantom{+}0 \\0&\hphantom{+}0 & \hphantom{+}1 &\hphantom{+}1 \\0&\hphantom{+}0&\hphantom{+}1&-1\end{bmatrix} \]

If $m=1$, the basis is an expanded Hadamard gate and a SWAP gate:

\[ \textit{SWAP} \times (I \otimes H) = \frac{1}{\sqrt{2}}\begin{bmatrix}1& \hphantom{+}1 &\hphantom{+}0 &\hphantom{+}0\\0 &\hphantom{+}0 &\hphantom{+}1 &\hphantom{+}1 \\1&-1 & \hphantom{+}0 &\hphantom{+}0 \\0&\hphantom{+}0&\hphantom{+}1&-1\end{bmatrix}. \]

Therefore, given each value of $m$ Bob can send each quantum register $\ket{\alpha(x_i)}$ through the appropriate sequence of gates. The output will be either $v_1, v_2, v_3$ or $v_4$ and Bob will return an $(a,b)$ corresponding to the $v$ each register returns. 

This has shown how to construct the measurement vectors for implementation using standard quantum gates.

\subsection{Step 6}\label{sec:6}
In the final step the bank receives a value $(a,b)$ corresponding to each index in $\mathcal{L}_{holder}$. The bank knows the classical value $x \in \{0,1\}^4$ corresponding to each index position and the given $m \in \{0,1\}$. So the bank uses Def.~\ref{def:HMP} to verify whether $ \forall i \in \mathcal{L}_{holder} \; (x_i,m_i,a_i,b_i) \in \textit{HMP}_4$ and thus verify the coin. This is a classical step and involves checking whether each value $a$ sent by Bob gives the specified $b$ output and thus Bob's provided $(a,b)$ pairs satisfy the $\textit{HMP}_4$ condition. 

\new{
\section{Summary remarks}
\label{sec:summary}
We briefly review properties of the quantum coin scheme. Many of these arise from its basic design, rather than our implementation.}
\new{
\paragraph{Security} To compromise a coin an adversary must supply the correct pair $(a_i,b_i)$ corresponding to the $m=0$ or $1$ chosen by the bank for each register $i$. Gavinsky shows that makes the coins exponentially hard to counterfeit \cite{gavinsky2012quantum}. In addition, the coins cannot be directly cloned due to the quantum no-cloning theorem and can not be eavesdropped without collapsing the wave functions. Furthermore, unlike classical digital cash schemes, an intruder who pretends to be the bank cannot steal a valid coin from the holder. In fact, Gavinsky proves that this scheme is unconditionally secure even against adaptive multi-round attackers \cite{gavinsky2012quantum}.}
\new{
\paragraph{Efficiency} 
The number of verifications that a $\mathcal{Q}$-coin can go through is limited and there is a trade off between the size of the coin and the security of that coin. Gavinsky \cite{gavinsky2012quantum} shows that this dependence is optimal up to a polynomial.
}

\new{
The database of the bank is static, and therefore decentralised branches can exist which can perform verification. In addition the classical communication channel with the bank can remain unencrypted.
}

\new{
Our implementation uses common quantum gates. Other pairings or other gates could prove more efficient as the technology progresses. An integrated circuit of the gates could be used to improve the efficiency of the system.
}
\new{
\paragraph{Limitation}
One current limitation with quantum coins is the need for the quantum entanglement to persist for the lifetime of the coin. However, quantum entanglement is highly susceptible to decoherence, even tiny changes in its environment, such as atomic motion, can cause the entanglement to collapse. Most quantum entanglement technologies protect against decoherence from tens of $\upmu$s (micro-seconds) to tens of seconds. However Ion Traps show promise with \texttildelow12 days of entanglement lifetime. Unless a very high rate of coin turnover is envisaged then quantum memories will be required. Quantum memories are being developed \cite{yu2020entanglement}, but still have short-lifetimes, \texttildelow70$\upmu$s, as to not be practical at present.
}

\section{Conclusion}
\label{sec:conclusion}
In this paper, we have shown how to convert information stored on classical bits to information stored on quantum bits. We have demonstrated a quantum gate configuration that allows the implementation of Gavinsky's quantum coin scheme \cite{gavinsky2012quantum}. We have provided a brief outline of the code necessary for its deployment in SimulaQron \cite{dahlberg2018simulaqron}. We have also provided additional descriptions for various aspects of the protocol, building on the descriptions provided by Gavinsky. 

%% file: diagram.tex
\begin{center}
\small

    \begin{tikzpicture}[thick]
    %
    \tikzset{
operator/.style = {draw,fill=white,minimum size=1.5em},
operator2/.style = {draw,fill=white,minimum height=3cm},
phase/.style = {draw,fill,shape=circle,minimum size=5pt,inner sep=0pt},
surround/.style = {fill=white!10,thick,draw=black,rounded corners=2mm},
cross/.style={path picture={ 
\draw[thick,black](path picture bounding box.north) -- (path picture bounding box.south) (path picture bounding box.west) -- (path picture bounding box.east);
}},
crossx/.style={path picture={ 
\draw[thick,black,inner sep=0pt]
(path picture bounding box.south east) -- (path picture bounding box.north west) (path picture bounding box.south west) -- (path picture bounding box.north east);
}},
circlewc/.style={draw,circle,cross,minimum width=0.3 cm},
}
    %
    \node at (0,0) (q1) {\ket{0}};
    \node at (0,-1) (q2) {\ket{0}};
    %
    \node[operator] (op11) at (1,0) {H} edge [-] (q1);
    %
    \node[phase] (phase11) at (2,0) {} edge [-] (op11);
    \node[circlewc] (phase12) at (2,-1) {} edge [-] (q2);
    \draw[-] (phase11) -- (phase12);
    \node (end1) at (3,0) {} edge [-] (phase11);
    \node (end2) at (3,-1) {} edge [-] (phase12);
    %
    \draw[decorate,decoration={brace},thick] (3,0.2) to
    	node[midway,right] (bracket) {$\; = \frac{\ket{00}+\ket{11}}{\sqrt{2}} \equiv \begin{bmatrix} 1 &0 &0 &1\end{bmatrix}^T \myeq$ \ket{\Phi^+} }
	(3,-1.2);
    %
    %
    \end{tikzpicture}
\end{center}

%% file: diagram2.tex
\begin{center}
\footnotesize
    \begin{tikzpicture}[thick]
    %
    \tikzset{
operator/.style = {draw,fill=white,minimum size=1.5em},
operator2/.style = {draw,fill=white,minimum height=3cm},
phase/.style = {draw,fill,shape=circle,minimum size=5pt,inner sep=0pt},
surround/.style = {fill=white!10,thick,draw=black,rounded corners=2mm},
cross/.style={path picture={ 
\draw[thick,black](path picture bounding box.north) -- (path picture bounding box.south) (path picture bounding box.west) -- (path picture bounding box.east);
}},
crossx/.style={path picture={ 
\draw[thick,black,inner sep=0pt]
(path picture bounding box.south east) -- (path picture bounding box.north west) (path picture bounding box.south west) -- (path picture bounding box.north east);
}},
circlewc/.style={draw,circle,cross,minimum width=0.3 cm},
}
    %
    \node at (0,0) (q1) {\ket{0}};
    \node at (0,-1) (q2) {\ket{0}};
    %
    \node[operator] (op11) at (1,0) {H} edge [-] (q1);
    %
    \node[phase] (phase11) at (2,0) {} edge [-] (op11);
    \node[circlewc] (phase12) at (2,-1) {} edge [-] (q2);
    \draw[-] (phase11) -- (phase12);
    \node[operator] (h2) at (3,0) {H} edge [-] (op11);
    \node[operator] (i) at (3,-1) {I} edge [-] (q2);
    \draw[-] (h2) -- (i);
   
    \node (end1) at (4,0) {} edge [-] (h2);
    \node (end2) at (4,-1) {} edge [-] (i);
    %
    \draw[decorate,decoration={brace},thick] (4,0.2) to
	node[midway,right] (bracket) {$\; = \frac{\ket{00}+\ket{01}+\ket{10}-\ket{11}}{2}$ =\scriptsize{ $\begin{bmatrix}\hspace{-1em}\hphantom{+}1\hspace{-2em}&\hspace{-2em}\hphantom{+}1\hspace{-2em}&\hspace{-2em}\hphantom{+}1\hspace{-2em}&\hspace{-1em}-1\end{bmatrix}_A$}}
	(4,-1.2);
    \end{tikzpicture}
 \end{center}

%% file: main.bbl
\begin{thebibliography}{10}
\providecommand{\url}[1]{#1}
\csname url@samestyle\endcsname
\providecommand{\newblock}{\relax}
\providecommand{\bibinfo}[2]{#2}
\providecommand{\BIBentrySTDinterwordspacing}{\spaceskip=0pt\relax}
\providecommand{\BIBentryALTinterwordstretchfactor}{4}
\providecommand{\BIBentryALTinterwordspacing}{\spaceskip=\fontdimen2\font plus
\BIBentryALTinterwordstretchfactor\fontdimen3\font minus
  \fontdimen4\font\relax}
\providecommand{\BIBforeignlanguage}[2]{{%
\expandafter\ifx\csname l@#1\endcsname\relax
\typeout{** WARNING: IEEEtran.bst: No hyphenation pattern has been}%
\typeout{** loaded for the language `#1'. Using the pattern for}%
\typeout{** the default language instead.}%
\else
\language=\csname l@#1\endcsname
\fi
#2}}
\providecommand{\BIBdecl}{\relax}
\BIBdecl

\bibitem{miller2008quantum}
D.~A. Miller, \emph{Quantum mechanics for scientists and engineers}.\hskip 1em
  plus 0.5em minus 0.4em\relax Cambridge University Press, 2008.

\bibitem{benioff1980computer}
P.~Benioff, ``The computer as a physical system: A microscopic quantum
  mechanical {H}amiltonian model of computers as represented by {T}uring
  machines,'' \emph{J. statistical physics}, vol.~22, no.~5, pp. 563--591,
  1980.

\bibitem{igeta1988quantum}
K.~Igeta and Y.~Yamamoto, ``Quantum mechanical computers with single atom and
  photon fields,'' in \emph{International Quantum Electronics Conf.}\hskip 1em
  plus 0.5em minus 0.4em\relax Optical Society of America, 1988, p. TuI4.

\bibitem{ekert1991quantum}
A.~K. Ekert, ``Quantum cryptography based on {B}ell’s theorem,''
  \emph{Physical review letters}, vol.~67, no.~6, p. 661, 1991.

\bibitem{jones1998implementation}
J.~A. Jones and M.~Mosca, ``Implementation of a quantum algorithm on a nuclear
  magnetic resonance quantum computer,'' \emph{J. chemical physics}, vol. 109,
  no.~5, pp. 1648--1653, 1998.

\bibitem{marks2007quantum}
P.~Marks, ``Quantum cryptography to protect {S}wiss election,'' \emph{New
  Scientist}, 2007.

\bibitem{sasakione}
M.~Sasaki, ``One more step to commercialization: Study of quantum cryptographic
  network,'' \url{https://www.nict.go.jp/en/pdf/copy_of_NICT_NEWS_1102_E.pdf}.

\bibitem{qiu2014quantum}
J.~Qiu \emph{et~al.}, ``Quantum communications leap out of the lab.''
  \emph{Nature}, vol. 508, no. 7497, pp. 441--442, 2014.

\bibitem{markoff2015sorry}
J.~Markoff, ``{S}orry, {E}instein. {Q}uantum study suggests ‘spooky
  action’is real,'' \emph{The New York Times}, vol.~21, 2015.

\bibitem{vermaas2019quantum}
P.~Vermaas, D.~Nas, L.~Vandersypen, and D.~Elkouss~Coronas, ``Quantum internet:
  The internet's next big step,'' 2019.

\bibitem{mavroeidis2018impact}
V.~Mavroeidis, K.~Vishi, M.~D. Zych, and A.~J{\o}sang, ``The impact of quantum
  computing on present cryptography,'' \emph{arXiv preprint arXiv:1804.00200},
  2018.

\bibitem{shor1994algorithms}
P.~W. Shor, ``Algorithms for quantum computation: discrete logarithms and
  factoring,'' in \emph{Proc. 35\textsuperscript{th} Symposium on Foundations
  of Computer Science}.\hskip 1em plus 0.5em minus 0.4em\relax {IEEE}, 1994,
  pp. 124--134.

\bibitem{grover1996algorithms}
\BIBentryALTinterwordspacing
L.~K. Grover, ``A fast quantum mechanical algorithm for database search,'' in
  \emph{Proc. 28\textsuperscript{th} ACM Symposium on Theory of Computing},
  ser. STOC ’96.\hskip 1em plus 0.5em minus 0.4em\relax New York, NY, USA:
  Association for Computing Machinery, 1996, p. 212–219. [Online]. Available:
  \url{https://doi.org/10.1145/237814.237866}
\BIBentrySTDinterwordspacing

\bibitem{ripe}
{RIPE Network Coordination Centre}, \url{https://www.ripe.net/}.

\bibitem{github_rep_hackathon}
``{Quantum Internet Hackathon},'' \url{https://github.com/PEQI19/}, Nov. 2019.

\bibitem{connect_hosts_hackathon}
``{Connect hosts pan-European Quantum Internet Hackathon},''
  \url{https://www.techcentral.ie/connect-hosts-pan-european-quantum-internet-hackathon/},
  Nov. 2019.

\bibitem{gavinsky2012quantum}
D.~Gavinsky, ``Quantum money with classical verification,'' in \emph{2012 IEEE
  27\textsuperscript{th} Conference on Computational Complexity}.\hskip 1em
  plus 0.5em minus 0.4em\relax {IEEE}, 2012, pp. 42--52.

\bibitem{heisenberg1985anschaulichen}
W.~Heisenberg, ``{\"U}ber den anschaulichen inhalt der quantentheoretischen
  kinematik und mechanik,'' in \emph{Zeitschrift fur Physik 43}.\hskip 1em plus
  0.5em minus 0.4em\relax Springer, 1927, pp. 478--504.

\bibitem{bruss2000approximate}
D.~Bruss, G.~D’Ariano, C.~Macchiavello, and M.~Sacchi, ``Approximate quantum
  cloning and the impossibility of superluminal information transfer,''
  \emph{Phys Rev A}, vol.~62, no.~6, p. 062302, 2000.

\bibitem{chaum1988untraceable}
D.~Chaum, A.~Fiat, and M.~Naor, ``Untraceable electronic cash,'' in \emph{Conf.
  Theory and Application of Cryptography}.\hskip 1em plus 0.5em minus
  0.4em\relax Springer, 1988, pp. 319--327.

\bibitem{wooters1982quantum}
W.~Wooters and W.~Zurek, ``Quantum no-cloning theorem,'' \emph{Nature}, vol.
  299, p. 802, 1982.

\bibitem{wiesner1983conjugate}
S.~Wiesner, ``Conjugate coding,'' \emph{ACM Sigact News}, vol.~15, no.~1, pp.
  78--88, 1983.

\bibitem{bennett1983quantum}
C.~H. Bennett, G.~Brassard, S.~Breidbart, and S.~Wiesner, ``Quantum
  cryptography, or unforgeable subway tokens,'' in \emph{Advances in
  Cryptology}.\hskip 1em plus 0.5em minus 0.4em\relax Springer, 1983, pp.
  267--275.

\bibitem{tokunaga2003anonymous}
Y.~Tokunaga, T.~Okamoto, and N.~Imoto, ``Anonymous quantum cash,'' in
  \emph{ERATO Conference on Quantum Information Science (EQIS)}, 2003.

\bibitem{mosca2010quantum}
M.~Mosca and D.~Stebila, ``Quantum coins,'' \emph{Error-Correcting Codes,
  Finite Geometries and Cryptography}, vol. 523, pp. 35--47, 2010.

\bibitem{lutomirski2010online}
A.~Lutomirski, ``An online attack against {W}iesner's quantum money,''
  \emph{arXiv preprint arXiv:1010.0256}, 2010.

\bibitem{nielsen2002quantum}
M.~A. Nielsen and I.~Chuang, \emph{Quantum computation and quantum
  information}.\hskip 1em plus 0.5em minus 0.4em\relax American Association of
  Physics Teachers, 2002.

\bibitem{bar2004exponential}
Z.~Bar-Yossef, T.~S. Jayram, and I.~Kerenidis, ``Exponential separation of
  quantum and classical one-way communication complexity,'' in \emph{Proc.
  36\textsuperscript{th} ACM symposium on Theory of computing}, 2004, pp.
  128--137.

\bibitem{garcia2011equivalent}
J.~C. Garcia-Escartin and P.~Chamorro-Posada, ``Equivalent quantum circuits,''
  \emph{arXiv preprint arXiv:1110.2998}, 2011.

\bibitem{dahlberg2018simulaqron}
A.~Dahlberg and S.~Wehner, ``Simulaqron — a simulator for developing quantum
  internet software,'' \emph{Quantum Science and Technology}, vol.~4, no.~1, p.
  015001, 2018.

\bibitem{github_rep_qcoin}
J.~F. Santos, H.~Murray, and J.~Horgan, ``{Quantum Coin GitHub Repository},''
  \url{https://github.com/facocalj/quantumcoin}, Nov. 2019.

\bibitem{yu2020entanglement}
Y.~Yu, F.~Ma, X.-Y. Luo, B.~Jing, P.-F. Sun, R.-Z. Fang, C.-W. Yang, H.~Liu,
  M.-Y. Zheng, X.-P. Xie \emph{et~al.}, ``Entanglement of two quantum memories
  via fibres over dozens of kilometres,'' \emph{Nature}, vol. 578, no. 7794,
  pp. 240--245, 2020.

\end{thebibliography}
